\documentclass[
 preprint,
 superscriptaddress,
 amsmath,amssymb,
 aps,
 prb,
10pt,
 twocolumn,
]{revtex4-2}

\usepackage{graphicx}
\usepackage{dcolumn}
\usepackage{bm}
\usepackage[dvipsnames]{xcolor}
\usepackage[colorlinks,citecolor=blue]{hyperref}

\begin{document}
\title{Local high chirality near exceptional points based on asymmetric backscattering}

\author{Jingnan Yang}
\affiliation{State Key Laboratory for Mesoscopic Physics and Frontiers Science Center for Nano-optoelectronics, School of Physics, Peking University, 100871 Beijing, China}
\author{Hancong Li}
\affiliation{State Key Laboratory for Mesoscopic Physics and Frontiers Science Center for Nano-optoelectronics, School of Physics, Peking University, 100871 Beijing, China}
\author{Sai Yan}
\affiliation{Beijing National Laboratory for Condensed Matter Physics, Institute of Physics, Chinese Academy of Sciences, Beijing 100190, China}
\author{Qihuang Gong}
\affiliation{State Key Laboratory for Mesoscopic Physics and Frontiers Science Center for Nano-optoelectronics, School of Physics, Peking University, 100871 Beijing, China}
\author{Xiulai Xu}
\email{xlxu@iphy.ac.cn}
\affiliation{State Key Laboratory for Mesoscopic Physics and Frontiers Science Center for Nano-optoelectronics, School of Physics, Peking University, 100871 Beijing, China}
\affiliation{Peking University Yangtze Delta Institute of Optoelectronics, Nantong, Jiangsu 226010, China}

\date{\today}

\begin{abstract}
We investigate local high chirality inside a microcavity near exceptional points (EPs) achieved via asymmetric backscattering by two internal weak scatterers. At EPs, coalescent eigenmodes exhibit position-dependent and symmetric high chirality characteristics for a large azimuthal angle between the two scatterers. However, asymmetric mode field features appear near EPs. Two azimuthal regions in the microcavity classified by the scatterers exhibit different wave types and chirality. Such local mode field features are attributed to the symmetries of backscattering in direction and spatial distribution. The connections between the wave types, the symmetry of mode field distribution and different symmetries of backscattering near EPs are also analyzed and discussed. Benefiting from the small size of weak scatterers, such microcavities with a high Q/V near EPs can be used to achieve circularly polarized quantum light sources and explore EP modified quantum optical effects in cavity quantum electrodynamics systems.
\end{abstract}
\maketitle

\section{\label{sec1}Introduction}
In recent years, optical microcavities have been widely used for studying non-Hermitian physics \cite{Heiss_2012, doi:10.1126/science.aar7709, PhysRevApplied.13.014070, PhysRevLett.130.143601, PhysRevApplied.19.034059}.
Exceptional points (EPs) are one of the most significant topics in non-Hermitian physics \cite{WDHeiss_2004, RevModPhys.93.015005, PhysRevLett.123.066405, Li2023}, at which eigenstates coalesce with identical eigenfrequencies.
For whispering galley mode (WGM) microcavities, two methods are usually used to achieve EPs. One is based on parity-time ($PT$) symmetry using balanced loss and gain \cite{Feng2017, El-Ganainy2018, PhysRevA.106.063526}.
The other is based on non-reciprocal interactions via asymmetric mode coupling induced by two nanoscatterers \cite{PhysRevA.84.063828, Hong2023} or a perfect reflection mirror \cite{Hashemi2022, PhysRevApplied.13.014070}.
So far, benefiting from the high-sensitivity to perturbations or chiral behaviors at EPs \cite{doi:10.1080/00018732.2021.1876991, Wiersig2020}, many applications have been reported, such as high-sensitivity detection \cite{Hodaei2017, Lai2019, PhysRevA.93.033809}, chiral lasing \cite{doi:10.1073/pnas.1603318113, Peng2014} and chiral absorption \cite{Soleymani2022, PhysRevLett.122.093901}.

Quantum light sources with specific polarization using single quantum emitters (QEs) are in great demand for applications in quantum information and quantum metrology \cite{RevModPhys.84.777, Wang2019}.
The single QE is usually weakly coupled with a microcavity to enhance the spontaneous emission rate for improving the brightness and indistinguishability of emitted single photons, and to modify the emission direction and polarization based on the Purcell effect \cite{https://doi.org/10.1002/lpor.201900425, Wang2019, Yang:21, Yang2023} and spin-momentum locking effect \cite{Lodahl2017, PhysRevB.101.205303, https://doi.org/10.1002/lpor.202100009}. Due to the nanoscale of a QE, its emission properties are locally decided by the electric field intensity and degree of circular polarization (DCP) where it's located instead of by global properties of cavity modes. Chiral optical modes at EPs have the potential of achieving circularly polarized single-photon sources \cite{PhysRevResearch.3.043096}. However, the application is limited by a low ratio of quality (Q) factor to mode volume (V) of the cavity at EPs. In fact, to exactly obtain EPs in experiment is nearly impossible which conflicts with the fact that EPs are sensitive to perturbations \cite{PhysRevApplied.12.024002}. Therefore, it's important to obtain a microcavity with a high Q/V and explore the mode field features near EPs, such as the distribution features of electric field and DCP.

Through designing and integrating two nanoscale holes as weak scatterers to induce controllable asymmetric backscattering, microcavities with both non-Hermitian designs and a high Q/V can be obtained \cite{PhysRevA.93.033809} for realizing EP enhanced circularly polarized single-photon emission or low-threshold chiral lasing \cite{Kim:14} and for studying light-matter interactions near EPs at the quantum level \cite{Chen2020, PhysRevLett.122.087401, PhysRevResearch.3.033029, PhysRevResearch.2.023375, PhysRevA.100.053841, PhysRevB.96.224303, PhysRevLett.120.213901, PhysRevLett.101.083601, Srinivasan2007}. Due to the modification of scatterers, the microcavity exhibits standing-wave features under symmetric backscattering or traveling-wave features at EPs with totally asymmetric backscattering, corresponding to orthogonality or coalescence of cavity modes. This indicates the connection between the wave types and the symmetry of backscattering in direction.

Here we study the mode field features of microcavities near EPs through numerical simulations. The EPs are achieved via asymmetric backscattering induced by two different weak scatterers integrated into a microdisk. Symmetrical mode field features in two azimuthal regions classified by the two scatterers and position-dependent high chirality up to $\pm1$ are observed at EPs when the two azimuthal regions have similar areas. Asymmetric mode field features are observed near EPs, where the two azimuthal regions exhibit different features in wave type and chirality. The asymmetric and local mode field features are induced by the asymmetry of backscattering in two azimuthal regions and the difference of decay rates for traveling and standing waves, and are differently affected by the scatterer parameters. This work reveals that the wave types and chirality inside the cavity are locally decided by the symmetry of backscattering in direction, while the asymmetry of backscattering in spatial distribution results in local and asymmetric mode field features.
\begin{figure*}
	\centering
	\includegraphics[scale=0.18]{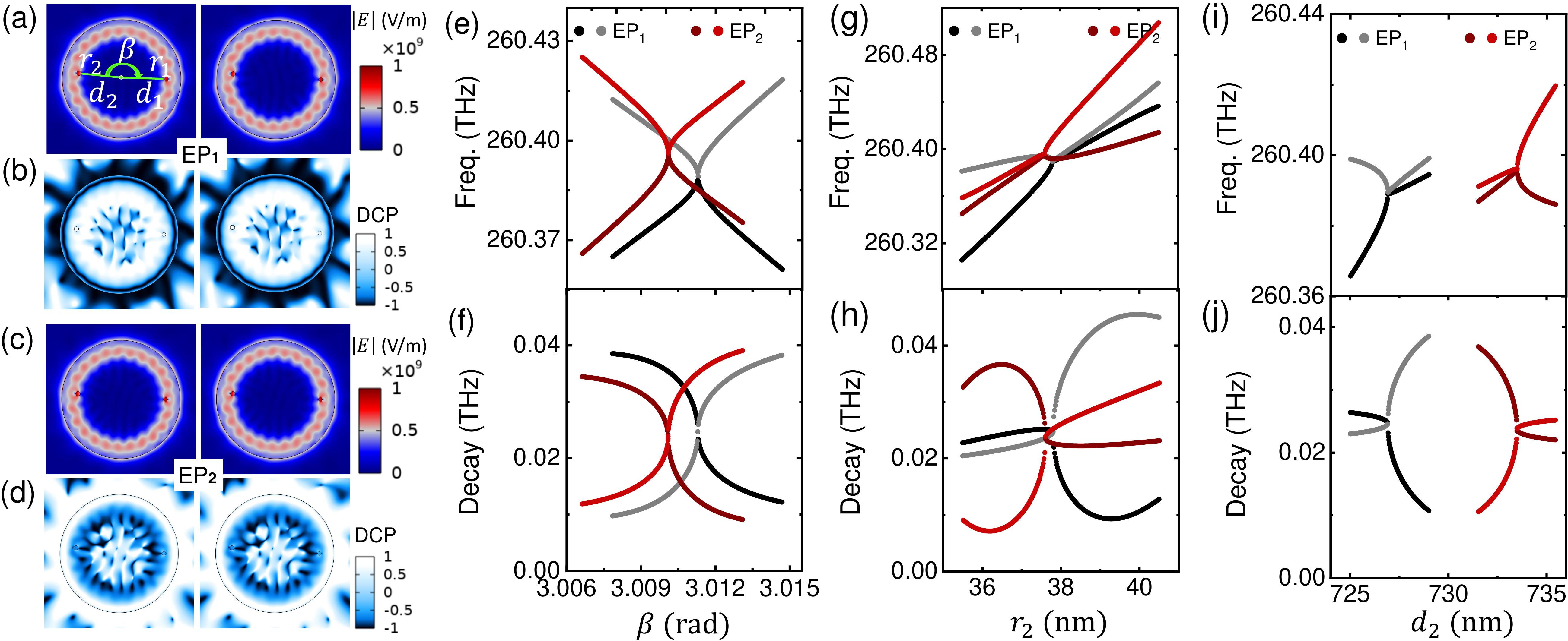}
	\caption{(a-b) Distributions of electric field $\vert E \vert$ (a) and DCP (b) of two modes at EP${_{1}}$. (c)-(d) Distributions of electric field $\vert E \vert$ (c) and DCP (d) of two modes at EP${_{2}}$. (e)-(j) Resonance frequencies and decay rates of two cavity modes near EP${_{1}}$ (black and gray) and EP${_{2}}$ (red) when changing $\beta$ (e)-(f), $r_{2}$ (g)-(h) and $d_{2}$ (i)-(j), respectively. The parameters for EP${_{1}}$ and EP${_{2}}$ are $r_{1}^{EP}\approx32.0$ nm, $d_{1}^{EP}\approx753.8$ nm, $r_{2}^{EP}\approx37.8$ nm, $d_{2}^{EP}\approx726.9$ nm and $\beta^{EP} \approx3.0112924$, and $r_{1}^{EP}\approx31.4$ nm, $d_{1}^{EP}\approx753.8$ nm, $r_{2}^{EP}\approx37.6$ nm, $d_{2}^{EP}\approx733.5$ nm and $\beta^{EP} \approx3.01009295$, respectively.}
	\label{p1}
\end{figure*}

\section{\label{sec2}Simulations and Design}
An unperturbed WGM microcavity has two degenerate cavity modes, propagating along clockwise (CW) or counter-clockwise (CCW) direction, respectively, namely, CW or CCW mode. Such a symmetric microcavity is very sensitive to perturbations like particles, which scatter the light and cause non-Hermitian mode coupling between CW and CCW modes. According to a simplified Hamiltonian picture of a microcavity perturbed by two weak scatterers \cite{PhysRevA.84.063828}, the backscattering of light from CW to CCW or from CCW to CW direction can be constructively canceled for proper scatterer parameters in size, location and refractive index, corresponding to two different EPs with coalescent CCW or CW modes and totally asymmetric backscattering, respectively.

 In principle, a larger overlap between the scatterer and the mode field usually causes a larger mode coupling and a large decay rate. To achieve EPs in a microcavity with high Q/V, we design two slightly different nanoscale air holes in size and location as weak scatterers with the mirror symmetry breaking in the microdisk. As shown in Fig. \ref{p1}(a), two air holes are designed into a microdisk with five parameters $r_{1}$, $d_{1}$, $r_{2}$, $d_{2}$ and $\beta$. $\beta$ is mathematically expected to satisfy $\beta \approx (N+1/2)\pi/m$ to cancel the scattering from CW to CCW or from CCW to CW direction depending on the backscattering parameters \cite{PhysRevA.84.063828}. $m$ is the azimuthal mode number. $N$ is an integral ($N<m$). We use two-dimensional simulation based on the finite element method. The microdisk with a radius of 1 $\mu$m is surrounded by an air layer and then a perfectly matched layer both with the depth larger than $\lambda/2$. The maximum mesh is 10 nm, about $\lambda/38$. The material is GaAs with an effective refractive index about 3.0377.

By changing the five parameters of the air holes, the mode coupling strength and the symmetry of backscattering in direction are controlled, as well as the eigenvalues and the wave types of cavity modes. We first consider a large $\beta$ with $N=11$ and $m=12$, where two azimuthal regions have similar areas. Through optimizing the parameters, as shown in Fig. \ref{p1} (a) and (c), two different EPs, labeled as EP${_{1}}$ and EP${_{2}}$, are obtained respectively, where coalescent modes exhibit traveling wave like features with vague nodes in the distribution of $\vert E \vert$. We then investigate the effects of $\beta$, $r_{2}$ and $d_{2}$ on the eigenvalues near EPs. The response of eigenvalues to $\beta$ is slightly asymmetric near EPs as shown in Fig. \ref{p1}(e)-(f), while that to $r_{2}$ or $d_{2}$ is highly asymmetric as shown in Fig. \ref{p1}(g)-(j). For example, the resonance frequencies (real parts of eigenvalues) and the decay rates (imaginary parts of eigenvalues) for $EP_1$ have similar values when $r_{2}>r_{2}^{EP}$ and $r_{2}<r_{2}^{EP}$, respectively. These eigenfrequency features are quite similar to those of coupled microcavities in the exact $PT$ and $PT$ broken phase as a function of the coupling strength $g$ between two microcavities \cite{Wang:23}.

\section{High chirality at EPs}
\begin{figure}
	\centering
	\includegraphics[scale=0.145]{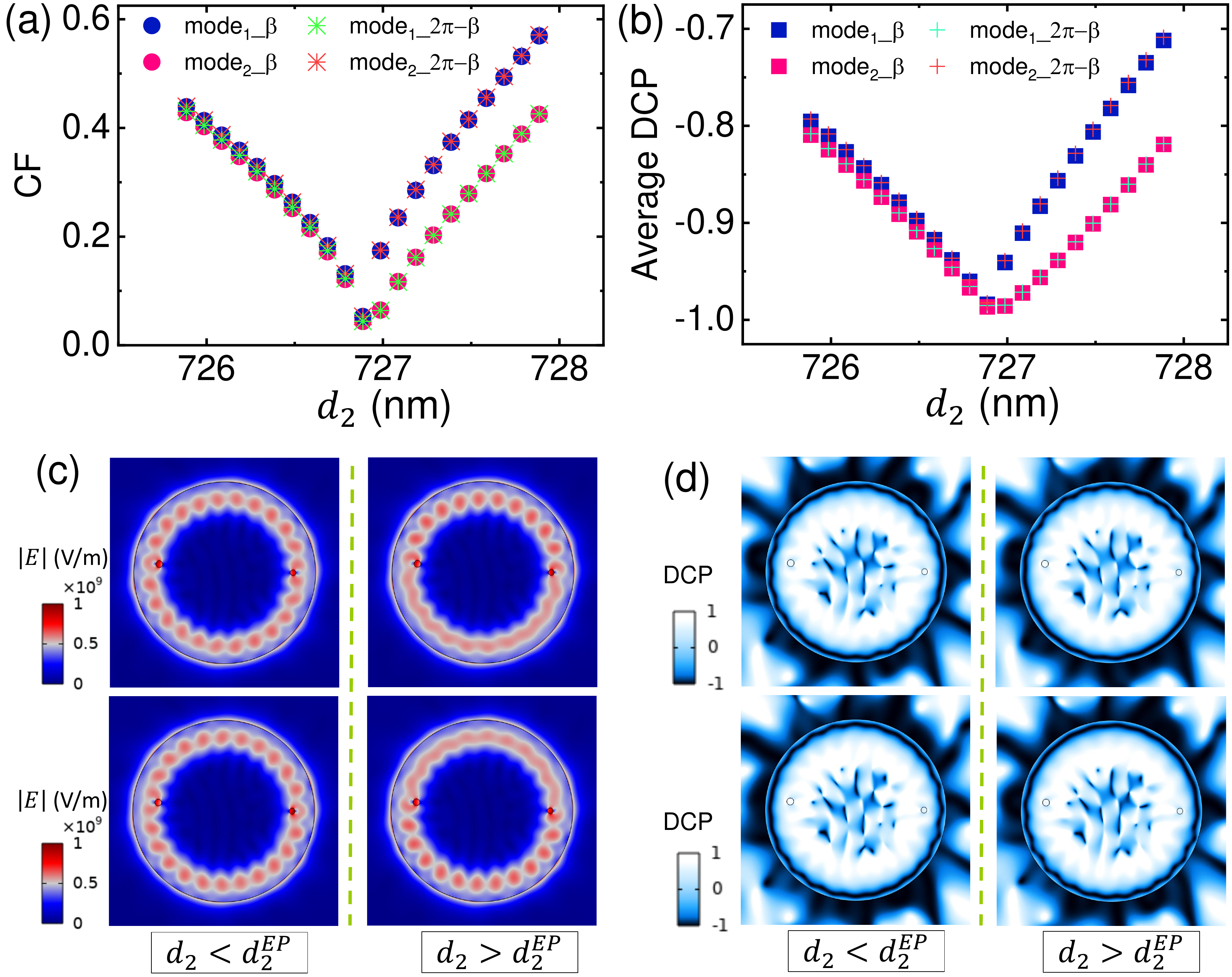}
	\caption{Local characteristics of electric field $\vert E \vert$ and DCP of cavity modes when scanning $d_2$ near $EP_1$. (a)-(b) Cost function (CF) (a) and average DCP (b) of the cavity modes distributed along a circle of $r_{detect}=820$ nm and a circle of $r_{detect}=940$ nm in azimuthal regions of $\beta$ and $2\pi-\beta$ when changing $d_2$, respectively. (c)-(d) Typical cases for $\vert E \vert$ (c) and DCP (d) distributions of two modes for $d_{2}<d_{2}^{EP}$ and $d_{2}>d_{2}^{EP}$, respectively.}
	\label{p2}
\end{figure}

Because of the nanoscale size of a QE, the DCP value locally determines the emission polarization of the QE. Therefore, we investigate the DCP distribution in a microcavity at EPs. The normalized DCP is defined as 2$\Im[ExEy^{*}]/(ExEx^{*}+EyEy^{*})$ \cite{PhysRevResearch.3.043096}, which describes the differences of $Ex$ and $Ey$ components in phase and amplitude. Here DCP of $\pm 1$ represents left/right circular polarization while DCP of $0$ represents linear polarization. As shown in Fig. \ref{p1}(b), two cavity modes exhibit identical DCP distributions at $EP_1$, where a ring-shape region in dark blue inside the microdisk and some outer region surrounding the microcavity exhibit DCP close to -1. Inwards, the DCP along a diameter of the microcavity decreases fast to zero and then increases almost up to 1 (see blue dots in Fig. \ref{p5}(f)). Outwards, the regions surrounding the microcavities exhibit high chirality, indicating the high chirality remains in the evanescent coupling using a waveguide. For $EP_2$, as shown in Fig. \ref{p1}(d), the cavity modes exhibit the opposite DCP characteristics. To be noted, some regions though exhibiting high chirality can be ignored because the mode field amplitude is too low. Here cavity modes at $EP_1$ and $EP_2$ exhibit relatively symmetric features in both $\vert E \vert$ and DCP distributions, which are different from EPs realized by two strong (larger size) scatterers with local and asymmetric mode field features \cite{PhysRevA.108.L041501}.

\section{Local mode field features near EPs}
\begin{figure}
	\centering
	\includegraphics[scale=0.145]{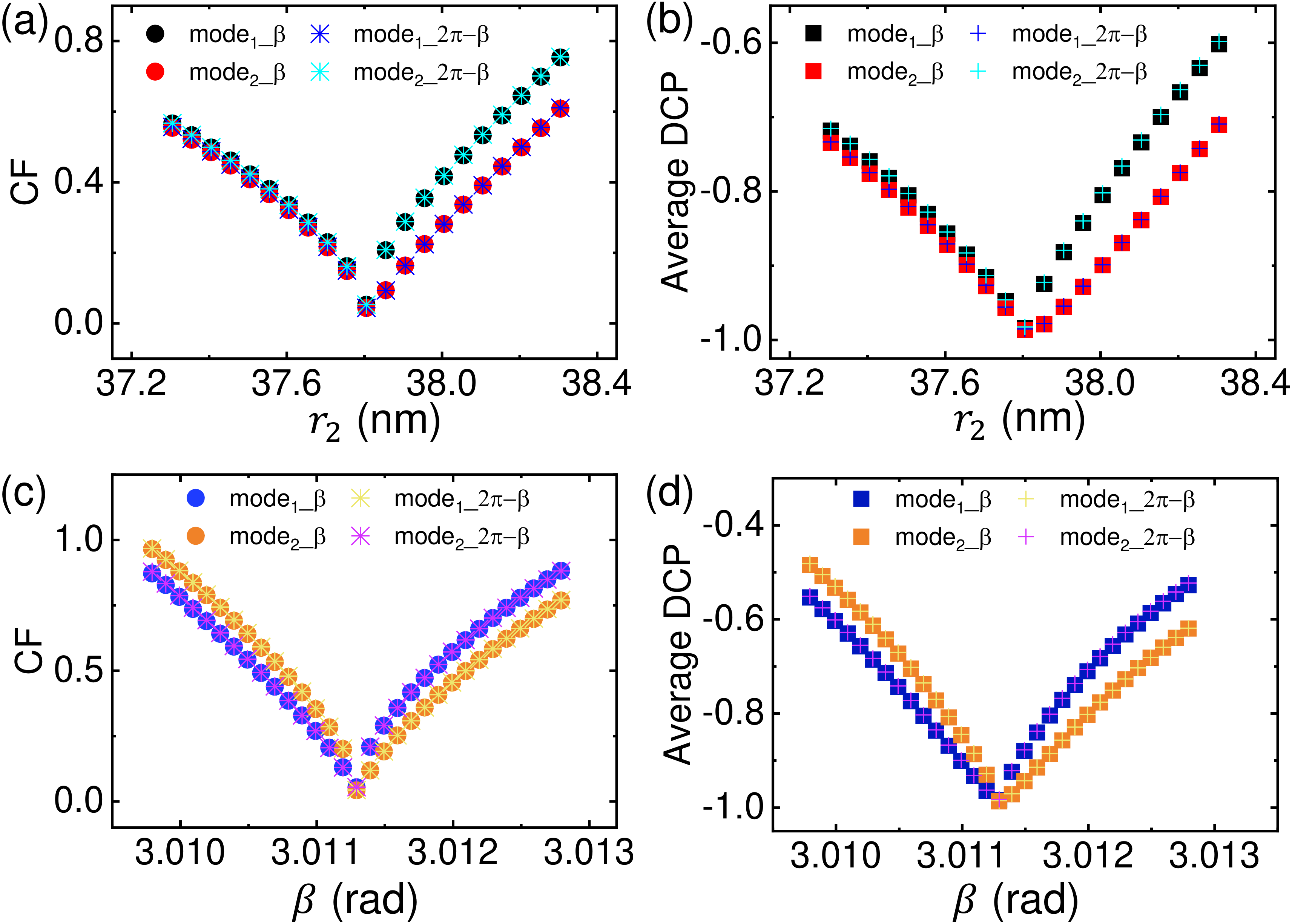}
	\caption{Local characteristics of $\vert E \vert$ and DCP of cavity modes distributed along a circle of $r_{detect}=820$ nm and a circle of $r_{detect}=940$ nm in different azimuthal regions when scanning $r_2$ and $\beta$ near $EP_1$, respectively. (a)-(b) CF (a) and average DCP (b) of cavity modes when changing $r_2$.  (c)-(d) CF (c) and average DCP (d) of cavity modes when changing $\beta$.}
	\label{p3}
\end{figure}

EPs are critical points in the parameter space, which are technically impossible to exactly reach in experiments. Therefore, we focus on the electric field $\vert E \vert$ and DCP distributions near EPs from the perspective of application. When analyzing the mode field distributions near $EP_1$, we find the mode field distributions are asymmetric in some cases, which are divided into two azimuthal regions of $\beta$ and $2\pi-\beta$ with different wave types and chirality by the two scatterers. To quantitatively describe such asymmetric features, we analyze $\vert E \vert$ and DCP of two modes distributed in the azimuthal regions of $\beta$ and $2\pi-\beta$ along a circle inside the microcavity with a radius $r_{detect}$ of $820$ nm and $940$ nm, respectively. Here $\vert E \vert$ and DCP have the maximum values at 820 nm and 940 nm at $EP_1$, respectively. When approaching EPs, the cavity modes will start exhibiting partly traveling wave features, where the difference between $\vert E \vert$ at antinode and node decreases. To quantify the wave type, we use the cost function (CF) based on the wave envelope and defined as  CF=$(\vert E \vert_{antinode}-\vert E \vert_{node})/(\vert E \vert_{antinode}+\vert E \vert_{node})$ \cite{MOTAHARIBIDGOLI2023117728}. CF$=0$ and CF$=1$ correspond to pure traveling waves and pure standing waves, respectively, while $0<$ CF $<1$ corresponds to the hybrid waves with a combination of both traveling and standing waves. $\vert E \vert_{antinode}$ and $\vert E \vert_{node}$ are obtained through sine fitting $\vert E \vert$ in the same azimuthal regions. To quantify the chirality, DCP values for azimuthal regions of $\beta$ and $2\pi-\beta$ are respectively averaged along the radians of $r_{detect}=940$ nm.
\begin{figure*}
	\centering
	\includegraphics[scale=0.19]{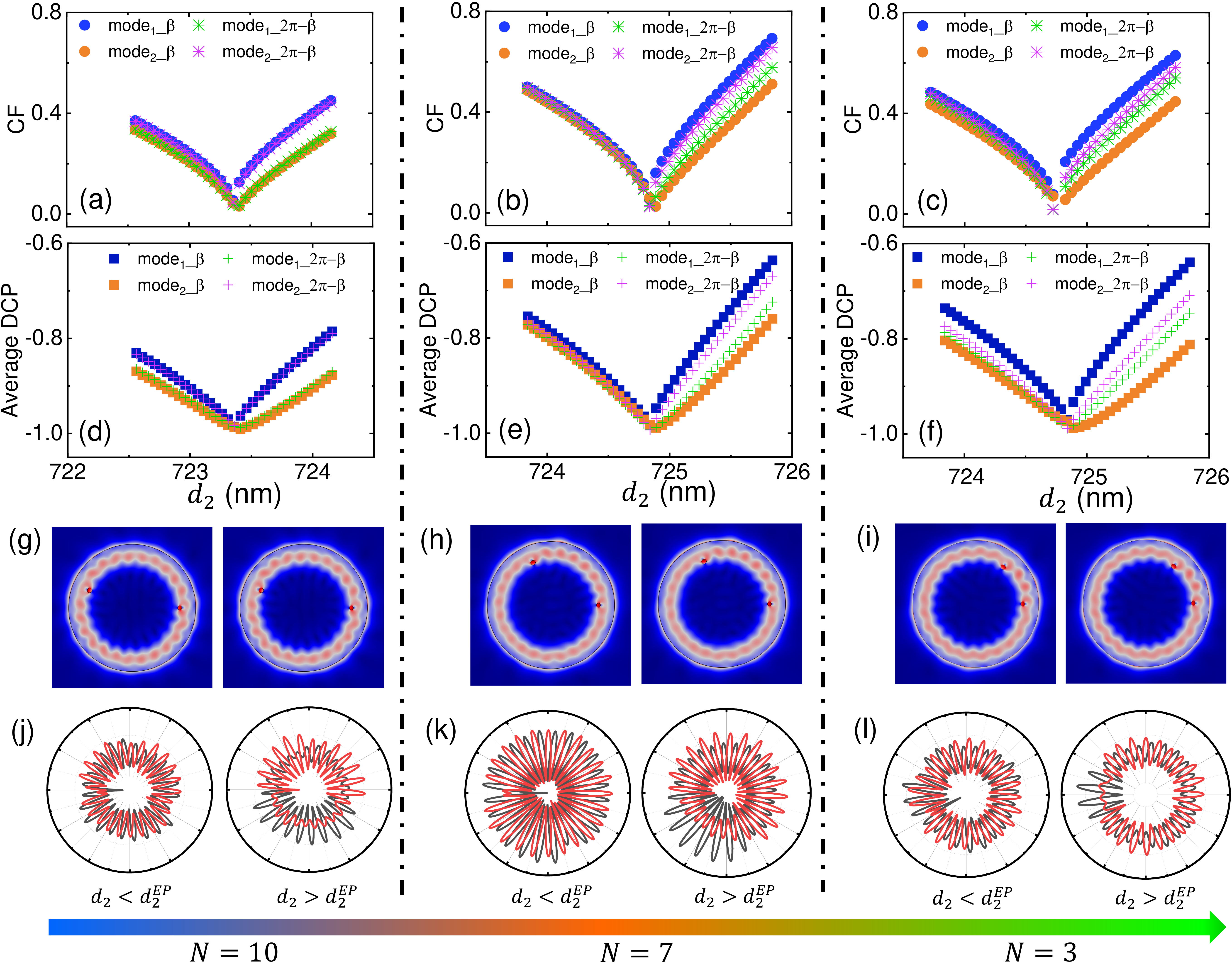}
	\caption{Local characteristics of $\vert E \vert$ and DCP of cavity modes when scanning $d_2$ near EPs with different $N$. (a)-(f) CF (a-c) and average DCP (d-f) of the cavity modes for $N=10, 7, 3$, respectively. (g)-(i) $\vert E \vert$ distributions of two cavity modes at EPs for $N=10, 7, 3$, respectively. (j)-(l) Polar maps of $\vert E \vert$ along a circle of $r_{detect}=820$ nm for $d_{2}<d_{2}^{EP}$ and $d_{2}>d_{2}^{EP}$.}
	\label{p4}
\end{figure*}

Figure \ref{p2}(a)-(b) show that CF (a) and average DCP (b) in two different azimuthal regions are symmetric for $d_2<d_{2}^{EP}$ but asymmetric for $d_2>d_{2}^{EP}$, where the differences of CF and DCP values between $\beta$ azimuthal region (circles) and $2\pi-\beta$ azimuthal region (stars) are very small (see mode field distribution in Fig. \ref{p2}(c) and (d) for $d_2<d_{2}^{EP}$), while CF and average DCP of both modes are different in $\beta$ azimuthal region and $2\pi-\beta$ azimuthal region for $d_2>d_{2}^{EP}$) (see mode field distribution in Fig. \ref{p2}(c) and (d) for $d_2>d_{2}^{EP}$). Specifically, DCP in Fig. \ref{p2} (d) is distributed similarly for $d_2<d_{2}^{EP}$ with the similar wave type, but distributed more smoothly only in $\beta$ or $2\pi-\beta$ azimuthal region with more traveling wave features for $d_2>d_{2}^{EP}$. Interestingly, the symmetry of mode field feature is closely related to the decay rates of the cavity modes. The cavity modes exhibiting symmetric mode field features usually have similar decay rates, while the cavity modes with $\beta$ ($\beta<2\pi-\beta$) azimuthal region exhibiting a higher CF (less traveling-wave like) always has a larger decay rate. This is because the traveling waves are a little worse confined compared to standing waves, resulting in a larger loss rate for the cavity mode with a lower CF in $2\pi-\beta$ azimuthal region. Usually the decay rates of two cavity modes are much more different for $d_2>d_{2}^{EP}$ compared to $d_2<d_{2}^{EP}$ as shown in Fig. \ref{p1}(g)-(j), which explains why the two cavity modes always exchange wave types in two azimuthal regions for $N=11$. Apparently, a higher CF usually corresponds to a lower $\vert$DCP$\vert$, through comparing CF and DCP in two azimuthal regions for $d_2>d_{2}^{EP}$ in Fig. \ref{p2}(a) and (b).

Wave types and chirality features when changing $r_2$ and $\beta$ are also studied. As shown in Fig. \ref{p3}(a) and (b), $r_2$ has similar effects on CF and average DCP to $d_2$, while cavity modes for both $\beta>\beta^{EP}$ and $\beta<\beta^{EP}$ exhibit slightly asymmetric mode field features as shown in Fig. \ref{p3}(c) and (d), which are consistent to the change of two decay rates in Fig. \ref{p1}(f) where the decay rates are slightly different for $\beta_{EP}<\beta$ and $\beta_{EP}>\beta$ compared to those for $d_2$ and  $r_2$ in Fig. \ref{p1}(h) and (j). Compared to the effects of $d_2$ and $r_2$, when change $\beta$ near $\beta^{EP}$, the areas of two azimuthal regions change oppositely. Therefore, the sign of the decay rate difference exchanges near $EP_1$ when changing $\beta$, and the CF and DCP exchange accordingly in different azimuthal regions. In a word, the asymmetric field features in a microcavity with two scatterers near EPs are closely related to the decay rates because traveling waves and standing waves decay differently in the microcavity.
\begin{figure*}
	\centering
	\includegraphics[scale=0.2]{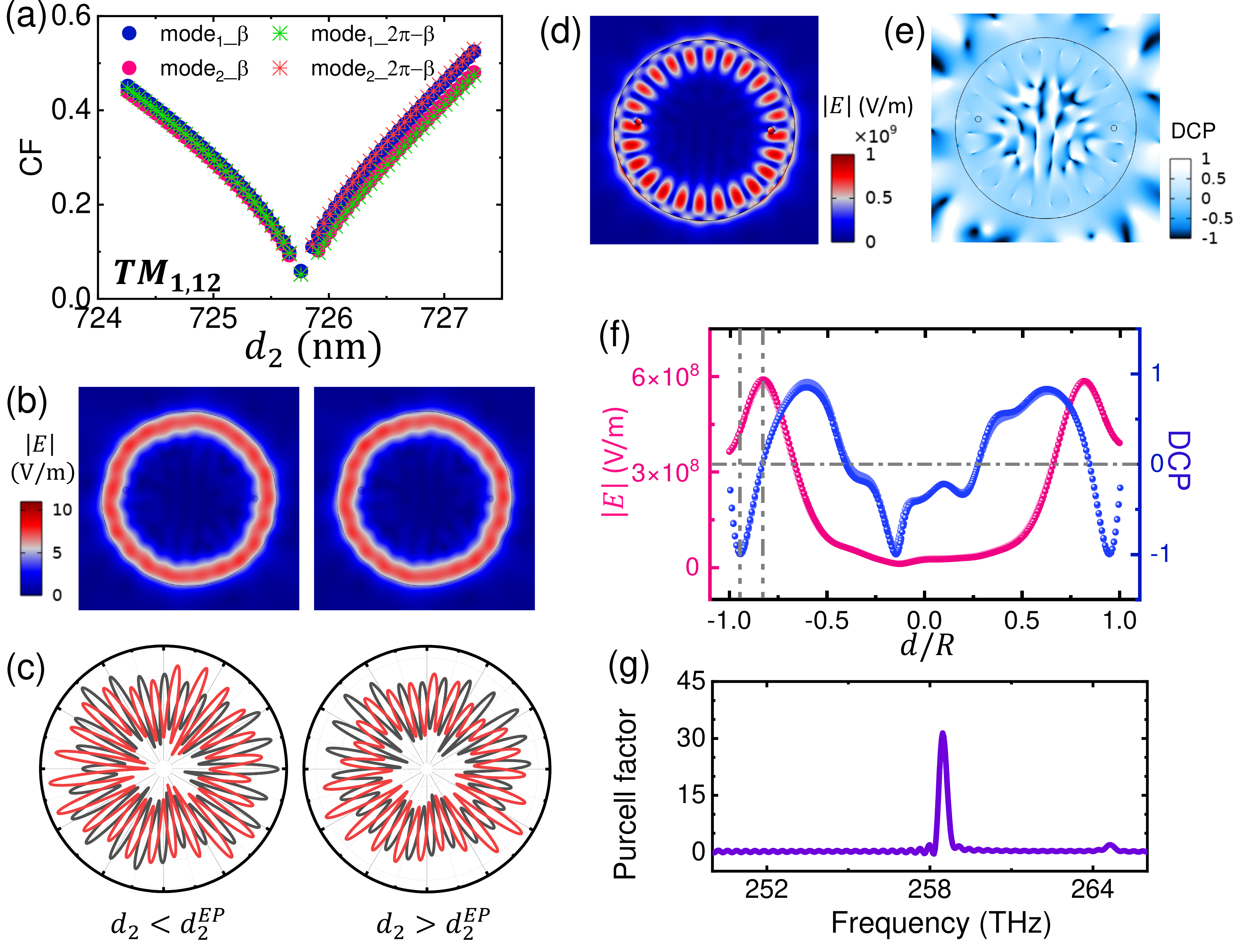}
	\caption{(a) Characteristics of CF of $TM_{1,12}$ modes near an EP with $N=11$ when scanning $d_2$. (b) $\vert E \vert$ distributions of $TM_{1,12}$ modes at EP. (c) Two typical polar maps of $\vert E \vert$ for $d_2<d_{2}^{EP}$ and $d_2>d_{2}^{EP}$ along a circle of $r_{detect}=820$ nm. (d)-(e) $\vert E \vert$ (d) and DCP (e) distributions of $TE_{1,12}$ mode of a microdisk with two identical air holes of $r_{1}=r_{2}\approx37.8$ nm, $d_{1}=d_{2}\approx726.9$ nm and $\beta\approx3.0112924$. (f) DCP and $\vert E \vert$ along a diameter inside the microcavity from Fig. \ref{p1}(a)-(b). (g) Purcell factor of a dipole in a microcavity near an EP at an antinode for the mode field.}
	\label{p5}
\end{figure*}

For $N=11$, the areas of two azimuthal regions are quite similar. To further explore the connections between the wave types, local features of mode field and the asymmetry of azimuthal region area. We consider $TE_{1,12}$ near EPs with different $N$. As shown in Fig. \ref{p4}(a)-(f), as $N$ decreases, the differences between CF at $\beta$ azimuthal region (circles) and $2\pi-\beta$ azimuthal region (stars) increase as well as DCP. The $\beta$ azimuthal region has a higher CF with more standing-wave like features, as shown in Fig. \ref{p4}(g)-(i). The small value of $\beta$ will directly increase the difference between CF in two azimuthal regions, due to the different scattering amplitudes of two scatterers in two azimuthal regions \cite{PhysRevA.108.L041501}. Such local features between two regions can also be observed for $d_2>d_{2}^{EP}$ when decreasing $N$. As shown in Fig. \ref{p4}(j)-(l), for $d_2>d_{2}^{EP}$, the differences of CF and DCP in $2\pi-\beta$ azimuthal region between two cavity modes decrease. For $d_2<d_{2}^{EP}$, the differences of CF and DCP between $\beta$ azimuthal region and $2\pi-\beta$ azimuthal region increase.

For comparison, the mode field distributions of $TM_{1,12}$ mode at an EP with $m=12$ and $N=11$ when scanning $d_2$ are also analyzed. As shown in Fig. \ref{p5}(a), asymmetric mode field features in wave type and chirality in two azimuthal regions are relatively less obvious than $TE_{1,12}$ mode in Fig. \ref{p2}(a). The polar maps of $\vert E \vert$ along a circle of $r_{detect}=820$ nm are relatively symmetric in two azimuthal regions for both $d_{2}<d_{2}^{EP}$ and $d_{2}>d_{2}^{EP}$ as shown in Fig. \ref{p5}(c). Compared to TE modes, TM modes are a little worse confined and distributed more approaching the microcavity edge as shown by $\vert E \vert$ distributions of coalescent cavity modes at EP in Fig. \ref{p5}(b). Therefore, the scatterers have a relatively smaller overlap with the mode field, which results in weaker backscattering and less asymmetric mode field features \cite{PhysRevA.108.L041501}. To further demonstrate that the high chirality of mode field results from EPs, $\vert E \vert$ and DCP distributions for a microcavity with two identical air holes are also calculated. The cavity modes exhibit obvious standing-wave-like features with distinguishable antinodes and nodes in mode field distributions as shown in Fig. \ref{p5}(d) and DCP with a low average value close to 0 as shown in Fig. \ref{p5}(e).

\section{Discussion}
 For $PT$-symmetry coupled microcavity systems, $PT$ broken phase corresponds to the case with identical resonance frequencies but different decay rates and mode fields distributed asymmetrically in two microcavities, indicating the $PT$ symmetry breaking of eigenstates. While $PT$ unbroken phase corresponds to the case with identical decay rates but different resonance frequencies and the mode field distributed symmetrically in two microcavities \cite{PhysRevLett.80.5243, El-Ganainy2018}.
 For the single microcavity with asymmetric backscattering in this work, eigenvalue features are similar to those in $PT$ symmetry and $PT$ broken phase as shown in Fig. \ref{p1}(g)-(j). Meanwhile, asymmetrical and local features are observed in spatial distribution of $\vert E \vert$ and DCP, similar to the spatial field distribution asymmetry in the $PT$ broken phase which are closely related to complex eigenvalues. The $PT$ symmetry breaking with the loss or the gain larger than the coupling strength $g$ is exhibited in asymmetric mode field intensity of two modes in two microcavities. For a microcavity with asymmetric backscattering, the symmetry breaking lies in the asymmetry of backscattering in two azimuthal regions divided by the two scatterers and is exhibited by asymmetric mode features in wave type and chirality in two different azimuthal regions. The dependence of the exchange of asymmetric mode field features in two azimuthal regions between two modes on the decay rates is because traveling waves decay faster than standing waves. Such eigenfrequency-related phenomena reveal the similarities and differences between non-Hermitian optical systems near EPs realized by different methods, and demonstrate that the wave types can be locally controlled in multiple-scatterer physical systems.

Through comparing the distributions of $\vert E \vert$ and DCP at EPs, it can be found that the maximum $\vert$DCP$\vert$ is mismatched with the maximum $\vert E \vert$. We extract the electric field amplitude $\vert E \vert$ and the DCP value of two modes along with a diameter of the microcavity. As shown in Fig. \ref{p5}(f), the maximum value of $\vert E \vert$ corresponds to DCP=0. The maximum value of $\vert$DCP$\vert$ corresponds to the location deviating outwards from the mode field antinode, where $\vert E \vert$ reduced to 2/3 of the maximum value but still can enable a high cavity enhancement of emission from a QE, carrying specific circular polarization. Both DCP and $\vert E \vert$ are position-dependent, therefore, a balance between them needs to be considered according to the expectations of obtaining the highest $\vert$DCP$\vert$ or the highest cavity enhancement for the QE. To further demonstrate the advantages of the high Q/V for our microcavity designs with two weak scatterers, Purcell factor of a dipole located at one antinode in mode field near an EP is calculated using three dimensional simulation based on finite-difference time-domain method. The microdisk is designed with a radius of 1 $\mu$m and a thickness of $220$ nm. As shown in Fig. \ref{p5}(g), a Purcell factor up to about 30 is obtained, which is high enough for applications in single-photon sources \cite{Senellart2017, Wang2019}. So far, EP enhanced emission of a single QE has been theoretically reported \cite{ Lu:22, PhysRevA.107.043714} but rarely experimentally reported, which is restricted by a large V \cite{Zhu:10} or a low Q factor \cite{PhysRevResearch.3.043096} for a microcavity at/near an EP. The microcavity with such non-Hermitian designs in this work can be used to explore EP modified quantum effects in cavity QED systems, such as Petermann effect \cite{Pick:17}, and to realize single-photon sources carrying specific polarization \cite{Yang2024}.

\section{Conclusion}
In summary, we study the symmetry of mode field features in wave types and chirality of cavity modes in a microcavity with asymmetric backscattering at/near EPs induced by two different weak nanoscatterers. Our results show that cavity modes at EPs exhibit very high chirality which is position-dependent. We also observe asymmetric mode field features in wave type and chirality of two azimuthal regions divided by the two scatterers near EPs. The asymmetry in the areas of two azimuthal regions will further strengthen the asymmetric mode field features while a weaker scattering can weaken the asymmetric features. The asymmetric field features near EPs are eigenfrequency related, attributed to the asymmetry of backscattering in two azimuthal regions and the different decay rates of traveling waves and standing waves. Furthermore, a Purcell factor up to 30 is obtained for the microdisk near an EP. Our work helps deeper understand how the symmetries of backscattering in direction and spatial distribution affect the wave type, chirality and the symmetry of mode field distribution in non-Hermitian multiple-scattering processes, and demonstrates that wave types and chirality can be locally controlled. Benefiting from the small size of air holes, microcavities with high Q/V near EPs can be obtained for studying EP modified quantum effects in cavity QED systems and realizing chiral quantum light sources with highly circular polarization.

\section{\label{sec6}Acknowledgments}
This work was supported by the National Key Research and Development Program of China (Grant No. 2021YFA1400700), the National Natural Science Foundation of China (Grants Nos. 62025507, 11934019, 92250301 and 12204020), China Postdoctoral Science Foundation (Grant No. 2022M710234).

\section{Competing interests} The authors declare no competing interests.

\end{document}